\documentclass[11pt]{article}
\usepackage{graphicx,amsfonts}
\newtheorem{thm}{T{\sc HEOREM}}[section]

\newcommand{\ol}[1]{\overline{#1}}
\newcommand{\mapright}[1]{%
   \smash{\mathop{%
   \hbox to 1cm{\rightarrowfill}}\limits^{#1}}}
\newcommand{\mapleft}[1]{%
   \smash{\mathop{%
   \hbox to 1cm{\leftarrowfill}}\limits^{#1}}}
\newcommand{\maplleft}[2]{%
   \smash{\mathop{%
   \hbox to 1cm{\leftarrowfill}}\limits_{#1}^{#2}}}

\begin{document}

\title{A geometric approach to singularity confinement and
algebraic entropy}
\author{Tomoyuki Takenawa}

\date{}
\maketitle
\begin{center}
{Graduate School of Mathematical Sciences, University of
Tokyo, Komaba 3-8-1, Meguro-ku, Tokyo 153-8914, Japan}\\

\end{center}
\begin{abstract}
A geometric approach to the equation found by Hietarinta and 
Viallet, which satisfies the singularity confinement criterion but 
exhibits chaotic behavior, is presented. It is shown that this equation 
can be lifted to an 
automorphism of a certain rational surface and 
can therefore be considered to be the action 
of an extended Weyl group of indefinite type. A method to calculate 
its algebraic entropy by using the theory of intersection numbers is 
presented.
\end{abstract}

\section{Introduction}

The singularity confinement method has been proposed
by Grammaticos, Ramani and Papageorgiou  \cite{grp} as a
criterion for the integrability of (finite or infinite dimensional) discrete dynamical
systems.  The singularity confinement method demands that
even if singularities would appear due to particular initial values,
such singularities have to disappear after a finite number of iteration steps
and that the information on the initial values can be recovered 
(hence the dynamical system
has to be invertible). 

However ``counter examples'' were found by Hietarinta and Viallet
\cite{hv}. These mappings satisfy the singularity confinement criterion,
but the orbits of their solutions exhibit chaotic behavior.
The authors of \cite{hv} introduced the notion of algebraic entropy in order to test the degree of complexity of successive iterations.
The algebraic entropy is defined as $s=\lim_{n\to \infty}\log(d_n)/n$
where $d_n$ is the degree of $n$th iterate.
This notion is linked with Arnold's complexity, since
the degree of a mapping gives the intersection number of the image
of a line and a hyperplane. While the degree grows exponentially for
a generic mapping, it was shown that it grows
only in the polynomial order for a large class of integrable mappings
\cite{hv,arnord,bv,otgr}.

Many discrete Painlev\'{e} equations were found by Ramani, Grammaticos,
Hietarinta, Jimbo and Sakai \cite{rgh,js} and have been extensively studied.
Recently it was shown by Sakai \cite{sakai} that all of these 
(from the point of view of symmetries) are
obtained by studying rational surfaces in connection with the extended
affine Weyl groups. Surfaces obtained by successive blow-ups \cite{hartshorne}
of ${\mathbb P}^2$ or
${\mathbb P}^1 \times {\mathbb P}^1$ have been studied by several authors in the 
theory of birational mappings with invariants of finite
($=m$ in the case of ${\mathbb P}^2$ and $=m-1$
in the case of ${\mathbb P}^1 \times {\mathbb P}^1$, $1\leq m\leq 8$)
point sets in a rational surface connected to the Weyl groups
\cite{cd,dolgachev,dolgachev2}. Looijenga \cite{looijenga} and
Sakai studied the case
of $m=9$, in which case the birational mappings are connected with the extended
affine Weyl groups and are obtained as Cremona transformations. Discrete
Painlev\'{e} equations are recovered as  particular cases.

Our aim in this letter is to characterize one of the mappings found by 
Hietarinta and Viallet from the
point of view of the theory of rational surfaces. 
As its space of initial values, 
we obtain a rational surface associated with some root system of 
indefinite type. 
Conversely we recover
the mapping from the surface and consequently obtain an extension of 
mapping to its non-autonomous version. 
By considering the intersection numbers of divisors, 
we also present a method to calculate the algebraic entropy of a mapping. 
It is shown that the degree of the mapping is given by the $n$th power 
of a matrix which is given by the action of the mapping on the Picard group.


\section{Construction of the space of initial values by blow-ups}

We consider the dynamical system written by the birational map
$\varphi:{\mathbb P}^1 \times {\mathbb P}^1 \to {\mathbb P}^1 \times {\mathbb P}^1 $
\begin{eqnarray}\label{hv}
(x_n,y_n)\mapsto(x_{n+1},y_{n+1})
=(y_n,-x_n+y_n+a/y_n^2)
\end{eqnarray}
where $a \in {\mathbb C}$ is a nonzero constant.
This equation was found by Hietarinta and Viallet \cite{hv} and we call
it the HV eq.. To test the singularity confinement,
let us assume $x_0\neq0$ and $y_0=\epsilon$ where $|\epsilon|\ll 1$.
With these initial values singularities appear at $n=1$ as $\epsilon \to 0$ 
and disappear
at $n=4$.
In this case the information on the 
initial values is hidden in the coefficients of higher degree $\epsilon$. 
However, taking suitable
rational functions of $x_n$ and $y_n$  we can find the information of the initial values as finite
values. The fact that the leading orders of
$(x_1^2 y_1-a)y_1,~ (x_2^3(y_2/x_2-1)^2-a)x_2$ and
$(x_3 y_3^2-a)x_3 $ become
$ - a x_0, -a x_0$ and $ - a x_0$ actually suggests that the HV eq. can be
lifted to an automorphism of a suitable rational surface. 
Although of course these rational functions are
not uniquely determined.


Let the coordinates of ${\mathbb P}^1 \times {\mathbb P}^1$ be
$(x,y),(x,1/y),(1/x,y)$
and $(1/x,1/y)$ and let $x=\infty$ denote $1/x=0$.
We consider the inverse mapping of the HV eq.
\begin{eqnarray}\label{ihv}
\varphi^{-1}&:&(x,y)\mapsto
(\overline{x},\overline{y})=
(-y+x+a/x^2,x)
\end{eqnarray}
where $(\overline{x},\overline{y})$ means the image of $(x,y)$ by the mapping.
This mapping has two indeterminate points:
$(x,y)= (0,\infty),~(\infty,\infty)$.
We denote blowing up at $(x,y)=(x_0,y_0)\in {\mathbb C}^2$:
by
\begin{eqnarray}\label{coblow}
 (x,y) \leftarrow (x-x_0,(y-y_0)/(x-x_0))\cup
((x-x_0)/(y-y_0), y-y_0) 
\end{eqnarray}
By blowing up at $(x,y)=(x_0,y_0)$, 
$(x-x_0)/(y-y_0)$ takes meaning at this point.

First we  blow-up at $(x,y)= (0,\infty)$ ,
$(x,1/y) \leftarrow (x,1/xy)\cup (xy,1/y),$
and denote the obtained surface by $X_0$. Then
$\varphi^{-1}$ is lifted to a rational mapping 
from $X_0$ to ${\mathbb P}^1 \times {\mathbb P}^1$.
For example, in the new coordinates $\varphi^{-1}$ is expressed as
\begin{eqnarray*} 
(u_1,v_1):=(x,1/xy) &\mapsto&
(\overline{x},\overline{y})=((-u_1 + u_1^3 v_1+av_1)/(u_1^2 v_1), u_1)\\
(u_2,v_2):=(xy,1/y) &\mapsto&
(\overline{x},\overline{y})=((-u_2^2 v_2  + u_2^3 v_2^3  +a)/(u_2^2 v_2^2),
u_2v_2)
\end{eqnarray*}
where $u_1=0$ and $v_2=0$. This maps the exceptional
 curve at $(x,y)=(0,\infty)$ almost to 
$(\overline{x},\overline{y})=(\infty,0)$ but
 has an indeterminate point on the exceptional curve:
$(u_1,v_1)=(0,0).$ Hence we have to blow-up again at this point.
In general it is known that, if there is a rational mapping $X \to Y$ where
$X$ and $Y$ are smooth projective algebraic varieties,
the procedure of blowing up can be completed in a finite number of steps, after 
which one obtains a smooth projective algebraic variety $X_1$
such that the rational mapping is lifted to
a regular mapping from $X_1$ to $Y$
(theorem of the elimination of indeterminacy \cite{hartshorne}).

Here we obtain the following sequence of blow-ups. (For simplicity 
we take only one coordinate of (\ref{coblow}).) 
\begin{eqnarray*}
(x,y) &\maplleft{E_5}{\mbox{at}~(0,\infty)}& \left(x,\frac{1}{xy}\right)
\maplleft{E_6}{(0,0)} \left(x^2 y,\frac{1}{xy}\right) \\
&\maplleft{E_7}{(a,0)}& \left(xy(x^2 y-a),\frac{1}{xy}\right)
\maplleft{E_8}{(0,0)} \left(x^2y^2(x^2 y-a),\frac{1}{xy}\right)\\
(x,y) &\maplleft{E_9}{(\infty,\infty)}& \left(\frac{1}{x},\frac{x}{y}\right)
\maplleft{E_{10}}{(0,1)} \left(\frac{1}{x},x(\frac{x}{y}-1)\right)
\end{eqnarray*}
where the $E_i$ mean the total transforms generated by the blow-ups.
Of course the sequence above is not unique, since there is freedom to choose the coordinates.


We have obtained a mapping from $X_1$ to
${\mathbb P}^1 \times {\mathbb P}^1$ which is lifted from $\varphi^{-1}$.
But our aim is to construct a rational surface $X$ such that
$\varphi^{-1}$ is lifted to an automorphism of $X$. If it can be achieved,
$X$ is considered to be the space of initial values in the sense of Okamoto \cite{okamoto},
where a sequence of rational surfaces $X_i$ 
is (or $X_i$ themselves are) called the space of initial values
for a sequence of rational mappings $\varphi_i$ if  
each $\varphi_i$ is lifted to an isomorphism from $X_i$ to $X_{i+1}$ 
for all $i$.

First we construct the rational surface $X_2$ such that
$\varphi^{-1}$ is lifted to a regular mapping from $X_2$ to $X_1.$
For this purpose it is sufficient to eliminate the indeterminacy
of mapping from $X_1$ to $X_1$.
Consequently we obtain $X_2$ defined by the following sequence of blow-ups.
\begin{eqnarray*}
\left(\frac{1}{x},x(\frac{x}{y}-1)\right)
&\maplleft{E_{11}}{(0,0)}& \left(\frac{1}{x^2(x/y-1)},  x(\frac{x}{y}-1)\right)
\maplleft{E_{12}}{(0,0)} 
\left(\frac{1}{x^2(x/y-1)},  x^3(\frac{x}{y}-1)^2\right)\\
&\maplleft{E_{13}}{(0,a)}& \left(\frac{1}{x^2(x/y-1)},
 x^2(\frac{x}{y}-1)(x^3(\frac{x}{y}-1)^2-a)\right)\\
&\maplleft{E_{14}}{(0,0)}&
\left(\frac{1}{x^2(x/y-1)},x^4(\frac{x}{y}-1)^2(x^3(\frac{x}{y}-1)^2-a)\right)
\end{eqnarray*}

Next eliminating the indeterminacy of mapping from $X_2$ to $X_2$,
we obtain $X_3$ defined by the following sequence of blow-ups.
\begin{eqnarray*}
(x,y) &\maplleft{E_1}{(\infty,0)}& \left(\frac{1}{xy},y\right)
\maplleft{E_2}{(0,0)} \left(\frac{1}{xy},xy^2\right) \\
&\maplleft{E_3}{(0,a)}& \left(\frac{1}{xy}, xy(xy^2-a)\right)
\maplleft{E_4}{(0,0)} \left(\frac{1}{xy},x^2y^2(xy^2-a)\right)
\end{eqnarray*}

Here, it can be shown that the mapping from $X_3$ to $X_3$ which is lifted from $\varphi^{-1}$
does not have any indeterminate points and has an inverse 
(the mapping lifted from $\varphi$). Hence we obtain the
following theorem.
\begin{thm}
The HV eq. (\ref{hv}) can be lifted to an automorphism of $X (=X_3)$.
\end{thm}


\section{Action on the Picard group}

We denote the total transform of
$x={\rm constant},$ (or $y={\rm constant}$) on $X$ by $H_0$ (or $H_1$ resp.)
and the total transforms of the points subjected to blow-up
 by $E_1,E_2,\dots,E_{14}.$
It is known \cite{hartshorne} that the Picard group of $X$, Pic($X$),
is
$${\rm Pic}(X)= {\mathbb Z}H_0 + {\mathbb Z}H_1+  {\mathbb Z}E_1+\cdots+{\mathbb Z}E_{14}$$
and the canonical divisor of $X$, $K_X$, is
$$K_X=-2H_0 -2H_1 + E_1+\cdots+E_{14}$$
It is also known that the intersection numbers of $H_i$ or $E_k$ and
$H_j$ or $E_l$ are
\begin{eqnarray}\label{isn}
H_i \cdot H_j = 1-\delta_{i,j},~ E_k\cdot E_l= -\delta_{k,l},~
H_i \cdot E_k=0~
\end{eqnarray}
where $\delta_{i,j}$ is $1$ if $i=j$ and $0$ if $i\neq j.$

\begin{figure}[htbp]
\begin{center}
\includegraphics*[]{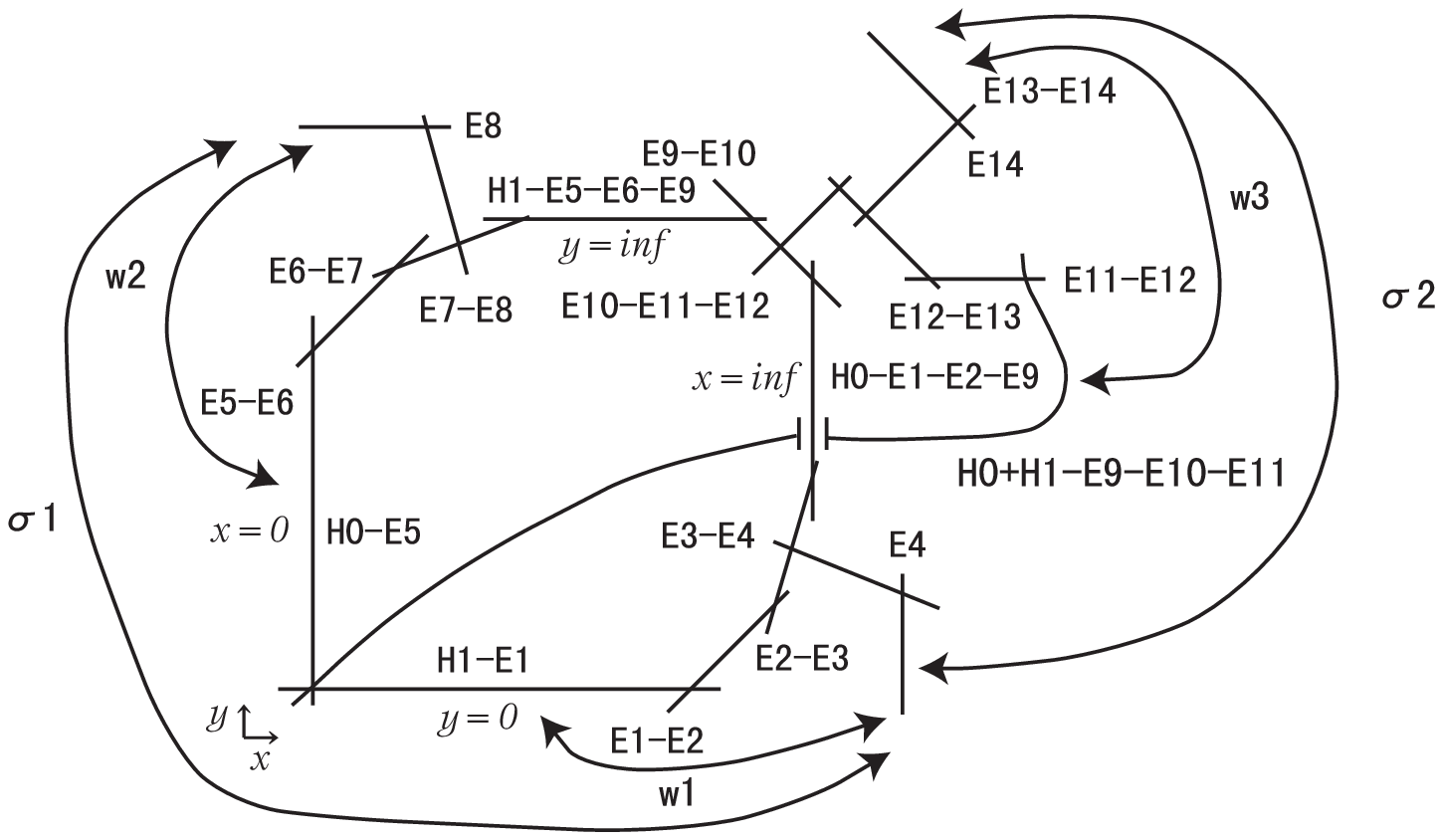}
\caption[]{}\label{divisors}
\end{center}
\end{figure}

We denote the proper transforms, i.e. prime divisors,
on $X$ by
$$H_0,H_1(\mbox{ 0 curve (self-intersection number is 0}))$$
$$C_0:=E_4,~C_1=E_8,~C_2:=E_{14},~C_3:=H_0-E_5,~C_4:=H_1-E_1 ~~(-1
\mbox{ curve)}$$
$$D_0:= E_1-E_2,~ D_1:= E_2-E_3,~ D_2:=E_3-E_4,~ D_3:=E_5-E_6,
~ D_4:=E_6-E_7,~D_5:=E_7-E_8,$$
$$D_6:=E_9-E_{10},~D_7:=E_{11}-E_{12},
~D_8:=E_{12}-E_{13}, ~D_9:=E_{13}-E_{14}~~(-2 \mbox{ curve)} $$
$$D_{10}:=H_0-E_1-E_2-E_9,~ D_{11}:=H_1-E_5-E_6-E_9,
~ D_{12}:=E_{10}-E_{11}-E_{12}~~(-3 \mbox{ curve)}$$
as in Fig.\ref{divisors}.
The intersection numbers of any pairs of divisors are given by
linear combinations of these divisors.

%
%
%

The anti-canonical divisor $-K_X$ can be reduced to the distinct
irreducible curves \\ $D_0,D_1,\cdots,D_{12}$ 
and the connection of $D_i$ are expressed by the Dynkin diagram:

\begin{picture}(400,100)
\put( 75,20){\line(1,0){20}}
\put(100,20){\line(1,0){20}}
\put(125,20){\line(1,0){20}}
\put(150,20){\line(1,0){20}}
\put(175,20){\line(1,0){20}}
\put(200,20){\line(1,0){20}}
\put(97.5,22.5){\line(0,1){20}}
\put(147.5,22.5){\line(0,1){20}}
\put(147.5,47.5){\line(0,1){20}}
\put(197.5,22.5){\line(0,1){20}}
\put(150,70){\line(1,0){20}}
\put(145,70){\line(-1,0){20}}
\put(72.5,20){\circle{5}}
\put(97.5,45){\circle{5}}
\put(97.5,20){\circle{5}}
\put(122.5,20){\circle*{5}}
\put(147.5,20){\circle{5}}
\put(172.5,20){\circle*{5}}
\put(197.5,20){\circle{5}}
\put(222.5,20){\circle{5}}

\put(147.5,45){\circle*{5}}
\put(147.5,70){\circle{5}}
\put(122.5,70){\circle{5}}
\put(172.5,70){\circle{5}}
\put(197.5,45){\circle{5}}
\put(72.5,10){\makebox(0,0){$D_3$}}
\put(97.5,10){\makebox(0,0){$D_4$}}
\put(122.5,10){\makebox(0,0){$D_{11}$}}
\put(147.5,10){\makebox(0,0){$D_6$}}
\put(172.5,10){\makebox(0,0){$D_{10}$}}
\put(197.5,10){\makebox(0,0){$D_{1}$}}
\put(222.5,10){\makebox(0,0){$D_0$}}
\put(112,45){\makebox(0,0){$D_5$}}
\put(162,45){\makebox(0,0){$D_{12}$}}
\put(212,45){\makebox(0,0){$D_2$}}
\put(122.5,80){\makebox(0,0){$D_9$}}
\put(147.5,80){\makebox(0,0){$D_8$}}
\put(172.5,80){\makebox(0,0){$D_7$}}

\put(280,50){\circle{5}}
\put(280,35){\circle*{5}}
\put(310,50){\makebox(0,0){$-2$ curve}}
\put(310,35){\makebox(0,0){$-3$ curve}}
\put(265,20){\line(1,0){20}}
\put(320,20){\makebox(0,0){intersection}}
\end{picture}

The HV eq.(\ref{hv}) acts on curves as
\begin{eqnarray}
\begin{array}{ll}
&(D_0,~D_1,~D_2,~D_3,~D_4,~D_5,~D_6,~D_7,~D_8,~
D_9,~D_{10},~D_{11},~D_{12},~C_0,~C_1,~C_2)\\
\to&
(D_5,~D_4,~D_3,~D_7,~D_8,~
D_9,~D_6,~D_0,~D_1,~ D_2,~
D_{11},~D_{12},~D_{10},~C_3,~C_2,~C_0)
\end{array}
\end{eqnarray}
and $C_4\mapsto C_1$. Hence the HV eq. acts on Pic($X$) as
\begin{eqnarray}\label{actp}
\left(
\begin{array}{c}
H_0\\
H_1,~~E_1,~~ E_2\\
E_3,~~E_4,~~E_5,~~E_6\\
E_7,~~E_8,~~E_9,~~E_{10}\\
E_{11},~~E_{12},~~E_{13},~~E_{14}
\end{array}\right)
\to
\left( \begin{array}{c}
3H_0+H_1-E_5-E_6-E_7-E_8-E_9-E_{10}\\
H_0,~~H_0-E_8,~~H_0-E_7 \\
H_0-E_6,~~H_0-E_5,~~E_{11},~~E_{12}\\
E_{13},~~E_{14},~~H_0-E_{10},~~H_0-E_9\\
E_1,~~ E_2,~~E_3,~~E_4
\end{array}\right)
\end{eqnarray}
(this table means $\overline{H_0}=3H_0+H_1-E_5-E_6-E_7-E_8-E_9-E_{10},$
$\overline{H_1}=H_0,$  $\overline{E_1}=H_0-E_8$ and so on.)
and their linear combinations. As we will see in section 6, the 
action (\ref{actp}) provides a method to calculate the
algebraic entropy of the HV eq.

\section{An extended Weyl group acting on the Picard group}

We shall decompose the action of the HV eq. on Pic($X$) as
a product of actions of order two elements of what turns out to be an extended
Weyl group.
Let us define the actions $\sigma_1,\sigma_2, w_1,w_2,w_3$ on Pic(X) as 
follows (See Fig.\ref{divisors}). (For simplicity we did not write 
the invariant elements under each action.) 
\begin{eqnarray}\label{weyld}
\begin{array}{lll}
&\sigma_1:&\left(\begin{array}{c}
H_0,~~H_1,~~E_1,~~E_2,~~E_3\\
E_4,~~E_5,~~E_6,~~E_7,~~E_8
\end{array}\right)
\to
\left( \begin{array}{c}
H_1,~~H_0,~~E_5,~~E_6,~~E_7\\
E_8,~~E_1,~~E_2,~~E_3,~~E_4
\end{array}\right)\\
&\sigma_2:&\left(\begin{array}{c}
H_1,~~E_1,~~E_2\\
E_3,~~E_4,~~E_9,~~E_{10}\\
E_{11},~~E_{12}~~E_{13},~~E_{14}
\end{array}\right)
\to
\left( \begin{array}{c}
H_0+H_1-E_9-E_{10},~~E_{11},~~E_{12}\\
E_{13},~~E_{14},~~H_0-E_{10},~~H_0-E_9\\
E_1,~~E_2,~~E_3,~~E_4
\end{array}\right)\\
&w_1:&\left(\begin{array}{c}
H_0,\\
E_1,~~E_2,~~E_3,~~E_4
\end{array}\right)
\to
\left(\begin{array}{c}
H_0+2H_1-E_1-E_2-E_3-E_4\\
H_1-E_4,~~H_1-E_3,~~H_1-E_2,~~H_1-E_1
\end{array}\right)\\
&w_2:&\left(\begin{array}{c}
H_1,\\
E_5,~~E_6,~~E_7,~~E_8
\end{array}\right)
\to
\left(\begin{array}{c}
2H_0+H_1-E_5-E_6-E_7-E_8\\
H_0-E_8,~~H_0-E_7,~~H_0-E_6,~~H_0-E_5
\end{array}\right)\\
&w_3:&\left(\begin{array}{c}
H_0,~H_1,~E_9,~E_{10}\\
E_{11},~E_{12},~E_{13},~E_{14}
\end{array}\right)
\to
\left(\begin{array}{c}
H_0+\alpha_3,~H_1+\alpha_3,~E_9+\alpha_3,~E_{10}+\alpha_3\\
E_{11}+\alpha_3^1,~E_{12}+\alpha_3^2,~E_{13}+\alpha_3^2,
~E_{14}+\alpha_3^1
\end{array}\right)
\end{array}
\end{eqnarray}
where $\alpha_3^1=H_0+H_1-E_9-E_{10}-E_{11}-E_{14}$,
$\alpha_3^2=H_0+H_1-E_9-E_{10}-E_{12}-E_{13}$ and
$\alpha_3=\alpha_3^1+\alpha_3^2$.

Then (\ref{actp}) becomes $w_2\circ \sigma_2 \circ \sigma_1$ and 
the following relations hold.
\begin{eqnarray}
&w_i^2=\sigma_j^2=1,~~(\sigma_1\sigma_2)^3=1,&\nonumber\\
&\sigma_1w_1=w_2\sigma_1,~~\sigma_1w_2=w_1\sigma_1,
~~\sigma_1w_3=w_3\sigma_1& \label{frel} \\
&\sigma_2w_1=w_3\sigma_2,~~\sigma_2w_2=w_2\sigma_2,~~
\sigma_2w_3=w_1\sigma_2&  \nonumber
\end{eqnarray}
\begin{flushleft}
\underline{The basis of the root system}\end{flushleft}
Let us define $\alpha_1,\alpha_2$ $\alpha_3\in {\rm Pic}(X)$ as
\begin{eqnarray}
\alpha_1&=&2H_1-E_1-E_2-E_3-E_4,\nonumber\\
\alpha_2&=&2H_0-E_5-E_6-E_7-E_8,\\
\alpha_3&=&2H_0+2H_1-2E_9-2E_{10}-E_{11}-E_{12}-E_{13}-E_{14}\nonumber
\end{eqnarray}
It is satisfied that $\alpha_i\cdot D_j=0$ for all $j$.
The actions of $\sigma_j$ and $w_k$ are
\begin{eqnarray}
&\sigma_j(\alpha_i) \mbox{ and } w_k(\alpha_i)&\nonumber\\
&\begin{array}{|c|c|c|c|c|c|}\hline
        &\sigma_1&\sigma_2&w_1&w_2&w_3\\ \hline
\alpha_1\mapsto&\alpha_2&\alpha_3&-\alpha_1&\alpha_1+2\alpha_2
             &\alpha_1+2\alpha_3\\\hline
\alpha_2\mapsto&\alpha_1&\alpha_2&\alpha_2+2\alpha_1&-\alpha_2
             &\alpha_2+2\alpha_3\\\hline
\alpha_3\mapsto&\alpha_3&\alpha_1&\alpha_3+2\alpha_1
             &\alpha_3+2\alpha_2&-\alpha_3\\\hline
\end{array}&
\end{eqnarray}
The action of $w_i$ can be written in the form
$w_i(\alpha_j)=\alpha_j-c_{ij}\alpha_i$ where 
$c_{ij}= 2(\alpha_i\cdot \alpha_j)/(\alpha_i\cdot \alpha_i)$.
Its Cartan matrix and Dynkin diagram are of the indefinite
type $H_{71}^{(3)}$ \cite{wan}:
\begin{eqnarray}\left[
\begin{array}{rrr}
2&-2&-2\\-2&2&-2\\-2&-2&2
\end{array}
\right]
\quad &\mbox{ and }&
\begin{picture}(0,25)(-20,30)
\put(-2.5,20){\circle{5}}
\put(-2.5,10){\makebox(0,0){$\alpha_1$}}
\put(42.5,20){\circle{5}}
\put(42.5,10){\makebox(0,0){$\alpha_2$}}
\put(20,42.5){\circle{5}}
\put(20,50){\makebox(0,0){$\alpha_3$}}
\put(2,19){\line(1,0){36}}
\put(2,21){\line(1,0){36}}
\put(1,24){\line(1,1){16}}
\put(-1,25){\line(1,1){16}}
\put(39,24){\line(-1,1){16}}
\put(41,25){\line(-1,1){16}}
\put(0,20){\line(2,1){10}}
\put(0,20){\line(2,-1){10}}
\put(-1,23){\line(1,6){2}}
\put(-1,23){\line(2,1){10}}
\put(17,41){\line(-1,0){10}}
\put(17,41){\line(-1,-6){2}}
\put(23,41){\line(1,0){10}}
\put(23,41){\line(1,-6){2}}
\put(40,20){\line(-2,-1){10}}
\put(40,20){\line(-2,1){10}}
\put(41,23){\line(-1,6){2}}
\put(41,23){\line(-2,1){10}}
\end{picture}
\end{eqnarray}

Hence it is seen \cite{kac} that the group of actions on Pic($X$) generated by ${w_i}$
and $\sigma_i$ coincides with the extended (including the full automorphism group
of the Dynkin diagram) Weyl group
of an indefinite type 
generated by
\begin{eqnarray}
<w_1,w_2,w_3,\sigma_1,\sigma_2> \label{coxe}
\end{eqnarray}
and the fundamental
relations (\ref{frel}).
From this fact we have the following theorem.
\begin{thm}
The HV eq. as the action of $w_2\sigma_2\sigma_1$ on Pic($X$) does not
commute with any element
of the group generated by $w_i$ and $\sigma_i$ except
$(w_2\sigma_2\sigma_1)^m$.
\end{thm}

\section{The inverse problem}
A birational mapping on a rational surface is called a 
Cremona transformation.
One method for obtaining a Cremona transformation (which exchanges a certain
pair of exceptional curves) is to interchange the blow down structures.
Following this method, we can construct the Cremona transformations
which yield the extended Weyl group (\ref{coxe}) and thereby recover the HV eq.
from its action on Pic($X$).  

These Cremona transformations are lifted to automorphisms of Pic($X$) but
do not have to be lifted to automorphisms of $X$ itself, i.e.
the blow-up points can be changed without however changing the 
intersection numbers (we consider isomorphisms from $X$ to $X'$,
where $X$ and $X'$ may have different blow-up points). 
In order to do this, one has to consider not 
only autonomous but also non-autonomous mappings.
By $a_0,a_1,a_2,a_3,a_4,a_5,a_6,a_7$, we denote the point where 
each $E_{10},E_3,E_4,E_7,E_8,E_{11},E_{13},E_{14}$ is generated by 
the blow-up (or its value of the coordinate).

Consequently, it can been seen that $w_2$ is written as $w_2=t \circ w_2'$,
where $w_2':(x,y)\mapsto (x,y-a_3/x^2-a_4/(a_3 x))$,
and $t$ is a certain automorphism of ${\mathbb P}^1\times{\mathbb P}^1.$
By taking suitable $t$, 
we can assume that the points of $1,5,9$-th blow-up are
fixed, $\overline{a_0}=a_0=1$ and $\overline{a_5}=a_5$. 
For the remaining points there are no such a priori requirements
and there evolution under the present isomorphism should be calculated
in detail. For example, under the above assumptions, $w_2$ can be seen to
reduce to
\begin{eqnarray}
w_2:&&(x,y;a_1,a_2,a_3,a_4,a_5,a_6,a_7)
\mapsto (\ol{x},\ol{y}~;~\ol{a_1},\ol{a_2},
\ol{a_3},\ol{a_4},\ol{a_5},\ol{a_6},\ol{a_7})\nonumber\\
&=&\left(x,y-\frac{a_3}{x^2}-\frac{a_4}{a_3x}~;~a_1,a_2- \frac{2a_1a_4}{a_3},
-a_3,a_4,a_5,a_6,a_7+\frac{2a_4a_6}{a_3}\right)
\end{eqnarray}
Here, in the calculation of the next iteration step we have to use 
$\overline{a_3}=-a_3$ instead of $a_3$.

Similarly $\sigma_1$ and $\sigma_2$ reduce to
\begin{eqnarray*}
\sigma_{1}:&&(x,y;a_1,a_2,a_3,a_4,a_5,a_6,a_7)\nonumber\\
&\mapsto&\left(-y,-x~;~-a_3,-a_4,
-a_1,-a_2,a_5,-a_6,a_7-4a_5^2a_6 \right)
\end{eqnarray*}
and
\begin{eqnarray*}
\sigma_{2}:&&(x,y;a_1,a_2,a_3,a_4,a_5,a_6,a_7)\nonumber\\
&\mapsto& \left(x,x-y-a_5~;~a_6,a_7- 2a_5^2a_6,
-a_3,a_4,a_5,a_1,a_2+2a_1a_5^2 \right)
\end{eqnarray*}

\begin{flushleft}
\underline{Non-autonomous HV eq.}
\end{flushleft}

The composition $w_2 \sigma_2 \sigma_1$ reduces to
\begin{eqnarray}
&&w_2 \circ \sigma_{2} \circ \sigma_{1}
:(x,y;a_1,a_2,a_3,a_4,a_5,a_6,a_7)
\mapsto (-y,x-y-a_5-\frac{a_1}{y^2}-\frac{a_2}{a_1y}~;~\nonumber\\
&& -a_6,a_7-2a_5^2a_6-\frac{2a_2a_6}{a_1},-a_1,-a_2,a_5,-a_3,
-a_4-2a_3a_5^2+\frac{2a_2a_3}{a_1})\label{nhv}
\end{eqnarray}
Of course this mapping satisfies the singularity confinement criterion
by construction and
in the case of $a_2=a_4=a_5=a_7=0$ and $a_1=a_3=a_6=a$ it 
coincides with the HV eq.(\ref{hv}) except their signs.
The difference between them comes from the assumption 
$\overline{a_5}=a_5$.
Assuming
$\overline{a_5}=-a_5$ by $w_2, \sigma_2$ and $\sigma_1$, 
we have another form of (\ref{nhv})
\begin{eqnarray}
&&w_2 \circ \sigma_{2} \circ \sigma_{1}
:(x,y;a_1,a_2,a_3,a_4,a_5,a_6,a_7)
\mapsto(y,-x+y+a_5+\frac{a_1}{y^2}+\frac{a_2}{a_1y}~;~\nonumber\\
&& a_6,-a_7+2a_5^2a_6+\frac{2a_2a_6}{a_1},a_1,a_2,-a_5,a_3,
a_4+2a_3a_5^2-\frac{2a_2a_3}{a_1})
\end{eqnarray}
Actually in the case of $a_2=a_4=a_5=a_7=0$ and $a_1=a_3=a_6=a$ 
it coincides with the HV eq.(\ref{hv}).


\section{Algebraic entropy}

In this section we consider the algebraic entropy which has been introduced
by Hietarinta and Viallet to describe the complexity of rational mappings 
\cite{hv}. 
The degree of a rational function 
$P(x,y)=f(x,y)/g(x,y)$, where 
$f(x,y)$ and $g(x,y)$ are polynomials and $P(x,y)$ is irreducible, 
is defined by
$$\deg(P)=\max\{\deg f(x,y),\deg g(x,y)\}$$
where $\deg(x^my^n)=m+n$.
The degree of the mapping $\varphi:(x,y)\mapsto(P(x,y),Q(x,y))$, where
$P(x,y)$ and $Q(x,y)$ are rational functions, is defined by
$$\deg(\varphi)=\max\{\deg P(x,y),\deg Q(x,y)\}$$
The algebraic entropy of the map $\varphi:(x,y)\mapsto(P(x,y),Q(x,y))$ 
is defined by 
$$\lim_{n\to\infty}\frac{1}{n}\log\deg(\varphi^n)$$
if the limit exists. The definition of algebraic entropy of $\varphi$ 
coincides with the definition for the case where $\varphi$
is a rational mappings from ${\mathbb P}^2 \to {\mathbb P}^2$.
It is known \cite{hv} that the algebraic entropy of the HV eq. becomes
$\log (3+\sqrt{5})/2$.
We recover the algebraic entropy of the HV eq.
by using the theory of intersection numbers. 

Let us define the curve $D$ in $X$ as $y/x=c$, where 
$c\in {\mathbb C}$ is a nonzero constant.
We denote the HV eq. by $\varphi$ and $(x_n,y_n)$ by 
$(P_n(x_0,y_0),Q_n(x_0,y_0))$.
By the fundamental theorem of algebra, 
$\deg_t(P_n(t,ct)) (=\deg(P_n(x,y))$ for $c\neq 1$) 
coincides with the intersection number of the curve
$x=P_n(t,ct)$ and the curve $x=d$ in ${\mathbb P}^1 \times {\mathbb P}^1$, where
$\deg_t(P(t))$ is the degree of the rational function of one variable $P(t)$ and $d\in {\mathbb C}$ is a nonzero constant.
It also coincides with the coefficient of $H_1$ as an element of Pic($X$). 
(Analogously for the intersection number of the curves related to $Q$ 
coincides with the coefficient of $H_0$.) 
The curve $D$ is expressed as $H_0+H_1-E_9$ in Pic($X$) if $c\neq 1$.   
Hence writing the coefficients of $H_0$ and $H_1$ of $\varphi^n(H_0+H_1-E_9)$ 
as $h_n^0,h_n^1$, 
we obtain the formula:
\begin{eqnarray*}
\deg(P_n)=h_n^1&
\deg(Q_n)=h_n^0
\end{eqnarray*}
The action of $\varphi$ on Pic($X$) is given by (\ref{actp}). 
Hence the algebraic entropy of the HV eq. becomes
$\lim_{n\to\infty}\frac{1}{n} \log\max\{h_n^0,h_n^1\}$.
In this case we have 
\begin{eqnarray*}
\log\max\{ |\mbox{ eigenvalues of } \varphi| \}&=&\log\frac{3+\sqrt{5}}{2}
\end{eqnarray*}

The proof of the theorems contained in this letter as well as
some other mappings, which satisfy the 
singularity confinement criterion
but which have positive algebraic entropy, will be discussed in a forthcoming
paper. \\
 
The author would like to thank H. Sakai, J. Satsuma, T. Tokihiro, A. Nobe, 
T. Tsuda, M. Eguchi and R. Willox 
for discussions and advice.


\clearpage

\end{document}